\documentstyle{article}
\date{}             
\begin{document}
\title{The $\pi NN$ and $\pi NN(1535)$ couplings in QCD }
\author{Shi-Lin Zhu$^1$, W-Y. P. Hwang$^2$ and Yuan-Ben Dai$^1$\\
$^1$Institute of Theoretical Physics, Academia Sinica\\ 
P.O.BOX 2735, Beijing 100080, P.R.China\\
and \\
$^2$Physics Department, National Taiwan University, Taipei}
\maketitle
\begin{center}
\begin{minipage}{120mm}
\vskip 0.6in
\begin{center}{\bf Abstract}\end{center}
{\large
We study the two point correlation function of two nucleon currents 
sandwiched between the vacuum and the pion state. 
The light cone QCD sum rules are derived for 
the $\pi NN$ coupling $g_{\pi NN} $ and 
the $\pi NN(1535)$ coupling $g_{\pi NN^\ast} $.
The contribution from the excited states and the continuum is subtracted 
cleanly through the double Borel transform with respect to the two external 
momenta, $p_1^2$, $p_2^2=(p-q)^2$.
We first improve the original sum rule for $g_{\pi NN} $ and 
determine the value of the pion wave function 
$\varphi_{\pi}(u_0 ={1\over 2})=1.5\pm 0.2$, which is a fundamental nonperturbative
quantity like the quark condensate.
Our calculation shows that the $\pi NN(1535)$ coupling is strongly suppressed.

\vskip 0.5 true cm
PACS Indices: 13.75.Gx; 14.20.Gk; 14.40.Aq; 13.75.Cs; 12.38.Lg
}
\end{minipage}
\end{center}

\large
%%%%%%%%%%%%%%%%%%%%%%%%%%%%%%%%%%%%%%%%%%%%%%%%%%%%%%%%%%%%%%%%%%%%%%%%
\section{Introduction}
\label{sec1}
%%%%%%%%%%%%%%%%%%%%%%%%%%%%%%%%%%%%%%%%%%%%%%%%%%%%%%%%%%%%%%%%%%%%%%%%
The $\pi NN$ coupling $g_{\pi NN} $ and the $\pi NN(1535)$ coupling 
$g_{\pi NN^\ast} $ play a very important role
in one boson exchange potentials (OBEP) for the nuclear force.
Although it is widely accepted that QCD is the underlying theory of 
the strong interaction, the self-interaction of the gluons causes 
the infrared behavior and the vacuum of QCD highly nontrivial. 
In the typical hadronic scale QCD is nonperturbative which makes 
the first principle calculation 
$g_{\pi NN} $ and $g_{\pi NN^\ast} $ unrealistic except the
lattice QCD approach, which is very computer time consuming. 
So a quantitative calculation of the $\pi NN$ and $\pi NN(1535)$ with 
a tractable and reliable theoretical approach proves valuable.

The method of QCD sum rules (QSR), as proposed
originally by Shifman, Vainshtein, and Zakharov \cite{SVZ} and adopted,
or extended, by many others \cite{RRY,IOFFE,BALIT}, are very 
useful in extracting the low-lying hadron masses and couplings.
In the QCD sum rule approach the nonperturbative QCD effects 
are partly taken into account through various condensates 
in the nontrivial QCD vacuum. In this work we shall use the light cone 
QCD sum rules (LCQSR) to calculate the $\pi NN$ and $\pi N N(1535)$ couplings.

The LCQSR is quite different from the conventional QSR, which is 
based on the short-distance operator product expansion. 
The LCQSR is based on the OPE on the light cone, 
which is the expansion over the twists of the operators. The main contribution
comes from the lowest twist operator. Matrix elements of nonlocal operators 
sandwiched between a hadronic state and the vacuum defines the hadron wave
functions. When the LCQSR is used to calculate the coupling constant, the 
double Borel transformation is always invoked so that the excited states and 
the continuum contribution can be treated quite nicely. Moreover, the final 
sum rule depends only on the value of the hadron wave function at a 
specific point, which is much better known than the whole wave function \cite{bely95}. 
In the present case our sum rules involve with the pion wave function (PWF)
$\varphi_{\pi}(u_0 ={1\over 2})$. Note this parameter 
is universal in all processes at a given scale.
In this respect, $\varphi_{\pi}(u_0 ={1\over 2})$ is a fundamental 
quantity like the quark condensate, which is to be determined 
with various nonperturbative methods. Like the quark condensate, it 
can be determined through the analysis of the light cone sum 
rules. In \cite{bely-z} the value is obtained as $\varphi_{\pi}(u_0 ={1\over 2})=1.2\pm 0.3$ 
using the pion nucleon coupling constant and the phenomenological $\rho\omega\pi$ coupling
constant as inputs. 

The LCQSR has been widely used to derive the couplings of pions with heavy mesons 
in full QCD \cite{bely95}, in the limit of $m_Q\to \infty$ 
\cite{zhu1} and $1/m_Q$ correction \cite{zhu3}, the couplings 
of pions with heavy baryons \cite{zhu2}, the $\rho$ decay widths of 
excited heavy mesons \cite{zhu4}, and various semileptonic decays of heavy 
mesons \cite{bagan98} etc.

The aim of the paper is two-fold. First we improve the original 
sum rule for $g_{\pi NN} (q^2=0)$ \cite{bely-z} in the following
way: (1) the contribution 
from the gluon condensate $<g_s^2 G^2>$ and the quark gluon mixed condensate 
$\langle g_c {\bar q}\sigma \cdot G q\rangle$ are calculated, 
which is numerically not negligible as in the case of the 
nucleon mass sum rule \cite{IOFFE} and the sum rule for the 
pion nucleon coupling constant \cite{zsl}; 
(2) the twist four contribution is collected in a more transparent form and 
is estimated slightly differently from \cite{bely-z};
(3) the uncertainty due to $\lambda_N$ is reduced in the numerical analysis 
with the help of the Ioffe's mass sum rule.
We arrive at $\varphi_\pi (1/2)=1.5\pm 0.2$ using the experimentally 
precisely known $g_{\pi NN}$ \cite{nijimegen}. 
Secondly we employ the LCQSR to calculate $\pi NN(1535)$ coupling. 
The continuum and excited states contribution is subtracted 
rather cleanly within our approach. Our result shows that the $\pi NN(1535)$
coupling is strongly suppressed. In both cases the final sum rules are 
stable with reasonable variations of the Borel parameter and 
the continuum threshold.

Our paper is organized as follows:
Section \ref{sec1} is an introduction.
We introduce the two point function for the $\pi NN$
vertex and saturate it with nucleon intermediate states in section \ref{sec2}. 
The definitions of the PWFs are presented in section \ref{sec3}.
In the following section we present the LCQSR for the $\pi NN$ coupling 
and discussions of these PWFs.
In section \ref{sec5} we employ LCQSR to calculate the $\pi NN(1535)$ coupling. 
A short summary is given in the last section.

%%%%%%%%%%%%%%%%%%%%%%%%%%%%%%%%%%%%%%%%%%%%%%%%%%%%%%%%%%%%%%%%%%%%%%%%
\section{Two Point Correlation Function for the $\pi NN$ coupling}
\label{sec2}
%%%%%%%%%%%%%%%%%%%%%%%%%%%%%%%%%%%%%%%%%%%%%%%%%%%%%%%%%%%%%%%%%%%%%%%%
The pion nucleon coupling constant $g_{\pi NN}$ has been calculated with the 
following variations of the traditional QCD sum rule method:
(1) the three point correlator of two nucleon and one pseudo-scalar
meson interpolating fields, which are saturated with resonances in the nucleon
$N$ and pion $\pi$ channels on the phenomenological side
\cite{RRY,RRY2,Rei,NP};
(2) the two point function of two nucleon interpolating fields sandwiched
between the vacuum and one $\pi$ state and saturating only with $N$
resonances \cite{RRY,Rei,SH};
(3) the external field method, which considers the two point 
correlator of two nucleon interpolating fields in the presence of the 
pion field \cite{zsl};
(4) the light cone QCD sum rules \cite{bely-z};
(5) using the Goldberger-Treiman relation with the nucleon axial 
coupling constant $g_A$ determined by QCD sum rules in external fields
\cite{jetp}. The recent calculation of $g_A=1.37\pm 0.10$ \cite{ioffe98} 
yields ${g^2_{\pi NN}\over 4\pi}=14.7\pm 2.1$.

In the method (1) the singular pole structure ${{\hat q}\gamma_5 \over q^2}$ 
was picked out and identified for the extraction of $g_{\pi NN} $. 
Note the operator product expansion (OPE) for the 
correlator (\ref{three-point}) is valid only in the region 
$p_1^2 \ll 0$, $p_2^2 \ll 0$, $q^2 \ll 0$. At $q^2 =0$ the OPE does 
not hold. Moreover it was pointed out that with the first methods 
the excited states and continuum contribution was not subtracted away 
\cite{IOFFE,maltman}, which could contaminate the sum rule severely.

We start with the two point function 
\begin{equation}
\Pi (p_1,p_2,q) = \int d^4 x e^{ip x} 
\left \langle 0 \vert {\cal T}
\eta_p (x)  {\bar{\eta_n}} (0) \vert \pi^+ (q) \right \rangle
\label{three-point}
\end{equation}
with $p_1 =p$, $p_2 = p-q$ and 
the Ioffe nucleon interpolating field \cite{IOFFE}
\begin{equation}\label{cur1}
\eta_p (x) = \epsilon_{abc}  [  u^a (x) {\cal C} \gamma_\mu
u^b (x) ] \gamma_5 \gamma^\mu d^c (x) \; ,
\label{eq4}
\end{equation}
\begin{equation}
{\bar\eta}_p(y) = \epsilon_{abc}[{\bar u}^b(y) \gamma_\nu C 
         {\bar u}^{aT}(y) ] {\bar d}^c(y) \gamma^\nu \gamma^5\; ,
\label{eq5}
\end{equation}
where $a,b,c$ is the color indices
and ${\cal C} = i \gamma_2 \gamma_0$ is the charge conjugation matrix.
For the neutron interpolating field, $u \leftrightarrow d$.

$\Pi (p_1,p_2,q)$ has the general form
\begin{equation}\label{eq3}
\Pi (p_1,p_2,q) =
 F (p_1 ^2 , p_2 ^2 ,q^2) {\hat q} \gamma_5 +
F_1 (p_1 ^2 , p_2 ^2 ,q^2) \gamma_5 + 
F_2 (p_1 ^2 , p_2 ^2 ,q^2) {\hat p} \gamma_5 +
F_3 (p_1 ^2 , p_2 ^2 ,q^2) \sigma_{\mu\nu}  \gamma_5 p^\mu q^\nu
\end{equation}

It was well known that the sum rules derived from the chiral odd tensor 
structure yield better results than those from the chiral even ones
in the QSR analysis of the nucleon mass and magnetic moment 
\cite{IOFFE,zsl-mag}. Most of the QSR analysis of 
the pion nucleon coupling constant deals with the tensor structure 
${\hat q} \gamma_5$. Based on these considerations we shall focus 
on the same tensor structure and study the function 
$F(p_1^2, p_2^2, q^2)$ in detail.

The pion nucleon coupling constant $g_{\pi NN}$ is defined by the
$\pi N$ interaction:
\begin{equation}
{\cal L}_{\pi NN} =  g_{\pi NN} {\bar N} i \gamma_5 {\bf \tau\cdot\pi} N 
+h.c.\; ,
\label{eq5a}
\end{equation}
where h.c. stands for the hermitian conjugate.

At the phenomenological level the eq.(\ref{eq2}) can be expressed as:
\begin{equation}\label{pole}
\Pi (p_1,p_2,q)  =   i \lambda_N^2 m_N g_{\pi NN} (q^2)
{  \gamma_5 {\hat q}
\over{ (p_1 ^2 - M_N ^2) (p_2 ^2 - M_N ^2) }} +\cdots
\label{phen1}
\end{equation}
where we include only the tensor structure $\gamma_5 {\hat q}$ only. 
The ellipse denotes the continuum and the single pole excited states to nucleon
transition contribution. $\lambda_N$ is the overlapping amplitude of 
the interpolating current $\eta_N (x)$ with the nucleon state
\begin{equation}
\left \langle 0 \vert \eta (0) \vert N (p) \right \rangle 
= \lambda_N  u_N (p)
\end{equation}

%%%%%%%%%%%%%%%%%%%%%%%%%%%%%%%%%%%%%%%%%%%%%%%%%%%%%%%%%%%%%%%%%%%%%%%%
\section{The Formalism of LCQSR and Pion Wave Functions}
\label{sec3}
%%%%%%%%%%%%%%%%%%%%%%%%%%%%%%%%%%%%%%%%%%%%%%%%%%%%%%%%%%%%%%%%%%%%%%%%
Neglecting the four particle component of the pion wave function, 	
the expression for $F(p^2_1,p^2_2,q^2)$ 
with the tensor structure at the quark level reads,
\begin{equation}\label{quark}
\int e^{ipx} dx
\langle 0| T\eta_p(x) {\bar \eta}_n(0) |\pi^+(q)\rangle =
-4{\rm i} \epsilon^{abc}
\epsilon^{a^{\prime} b^{\prime} c^{\prime}}  \gamma_{\mu} \gamma_5 S_d^{a a^{\prime}}(x)
\gamma_{\nu} C 
[\langle 0| u^b (x) {\bar d}^{b^{\prime}}(0)|\pi^+ (q)\rangle ]^T 
C \gamma_{\mu}
S_u^{c c^{\prime}}(x)\gamma_5 \gamma_{\nu}.
\end{equation}
where $iS(x)$ is the full light quark propagator with both perturbative  
term and contribution from vacuum fields. 
\begin{eqnarray}\label{prop}\nonumber
iS(x)=\langle 0 | T [q(x), {\bar q}(0)] |0\rangle &\\
=i{{\hat x}\over 2\pi^2 x^4} 
-{\langle {\bar q} q\rangle  \over 12}
-{x^2 \over 192}\langle {\bar q}g_s \sigma\cdot G q\rangle &\\ \nonumber
-{ig_s\over 16\pi^2}\int^1_0 du \{
{{\hat x}\over x^2} \sigma\cdot G(ux)-4iu {x_\mu\over x^2} 
G^{\mu\nu}(ux)\gamma_\nu \} +\cdots  & \; .
\end{eqnarray}
where we have introduced ${\hat x} \equiv x_\mu \gamma^\mu$.
The relevant feynman diagrams are presented in FIG 1. The squares 
denote the pion wave function (PWF). The broken solid line, broken curly line
and a broken solid line with a curly line attached in the middle
stands for the quark condensate, gluon condensate 
and quark gluon mixed condensate respectively. 

By the operator expansion on the light-cone
the matrix element of the nonlocal operators between the vacuum and 
pion state defines the two particle pion wave function.
Up to twist four the Dirac components of this wave function can be 
written as \cite{bely95}:
\begin{eqnarray}\label{phipi}
<\pi(q)| {\bar d} (x) \gamma_{\mu} \gamma_5 u(0) |0>&=&-i f_{\pi} q_{\mu} 
\int_0^1 du \; e^{iuqx} (\varphi_{\pi}(u) +x^2 g_1(u) + {\cal O}(x^4) ) 
\nonumber \\
&+& f_\pi \big( x_\mu - {x^2 q_\mu \over q x} \big) 
\int_0^1 du \; e^{iuqx}  g_2(u) \hskip 3 pt  , \label{ax} \\
<\pi(q)| {\bar d} (x) i \gamma_5 u(0) |0> &=& {f_{\pi} m_{\pi}^2 \over m_u+m_d} 
\int_0^1 du \; e^{iuqx} \varphi_P(u)  \hskip 3 pt ,
 \label{phip}  \\
<\pi(q)| {\bar d} (x) \sigma_{\mu \nu} \gamma_5 u(0) |0> &=&i(q_\mu x_\nu-q_\nu 
x_\mu)  {f_{\pi} m_{\pi}^2 \over 6 (m_u+m_d)} 
\int_0^1 du \; e^{iuqx} \varphi_\sigma(u)  \hskip 3 pt .
 \label{psigma}
\end{eqnarray}
\noindent 

\begin{eqnarray}
& &<\pi(q)| {\bar d} (x) \sigma_{\alpha \beta} \gamma_5 g_s 
G_{\mu \nu}(ux)u(0) |0>=
\nonumber \\ &&i f_{3 \pi}[(q_\mu q_\alpha g_{\nu \beta}-q_\nu q_\alpha g_{\mu \beta})
-(q_\mu q_\beta g_{\nu \alpha}-q_\nu q_\beta g_{\mu \alpha})]
\int {\cal D}\alpha_i \; 
\varphi_{3 \pi} (\alpha_i) e^{iqx(\alpha_1+v \alpha_3)} \;\;\; ,
\label{p3pi} 
\end{eqnarray}

\begin{eqnarray}
& &<\pi(q)| {\bar d} (x) \gamma_{\mu} \gamma_5 g_s 
G_{\alpha \beta}(vx)u(0) |0>=
\nonumber \\
&&f_{\pi} \Big[ q_{\beta} \Big( g_{\alpha \mu}-{x_{\alpha}q_{\mu} \over q \cdot 
x} \Big) -q_{\alpha} \Big( g_{\beta \mu}-{x_{\beta}q_{\mu} \over q \cdot x} 
\Big) \Big] \int {\cal{D}} \alpha_i \varphi_{\bot}(\alpha_i) 
e^{iqx(\alpha_1 +v \alpha_3)}\nonumber \\
&&+f_{\pi} {q_{\mu} \over q \cdot x } (q_{\alpha} x_{\beta}-q_{\beta} 
x_{\alpha}) \int {\cal{D}} \alpha_i \varphi_{\|} (\alpha_i) 
e^{iqx(\alpha_1 +v \alpha_3)} \hskip 3 pt  \label{gi} 
\end{eqnarray}
\noindent and
\begin{eqnarray}
& &<\pi(q)| {\bar d} (x) \gamma_{\mu}  g_s \tilde G_{\alpha \beta}(vx)u(0) |0>=
\nonumber \\
&&i f_{\pi} 
\Big[ q_{\beta} \Big( g_{\alpha \mu}-{x_{\alpha}q_{\mu} \over q \cdot 
x} \Big) -q_{\alpha} \Big( g_{\beta \mu}-{x_{\beta}q_{\mu} \over q \cdot x} 
\Big) \Big] \int {\cal{D}} \alpha_i \tilde \varphi_{\bot}(\alpha_i) 
e^{iqx(\alpha_1 +v \alpha_3)}\nonumber \\
&&+i f_{\pi} {q_{\mu} \over q \cdot x } (q_{\alpha} x_{\beta}-q_{\beta} 
x_{\alpha}) \int {\cal{D}} \alpha_i \tilde \varphi_{\|} (\alpha_i) 
e^{iqx(\alpha_1 +v \alpha_3)} \hskip 3 pt . \label{git} 
\end{eqnarray}
\noindent 
The operator $\tilde G_{\alpha \beta}$  is the dual of $G_{\alpha \beta}$:
$\tilde G_{\alpha \beta}= {1\over 2} \epsilon_{\alpha \beta \delta \rho} 
G^{\delta \rho} $; ${\cal{D}} \alpha_i$ is defined as 
${\cal{D}} \alpha_i =d \alpha_1 
d \alpha_2 d \alpha_3 \delta(1-\alpha_1 -\alpha_2 
-\alpha_3)$. 
Due to the choice of the
gauge  $x^\mu A_\mu(x) =0$, the path-ordered gauge factor
$P \exp\big(i g_s \int_0^1 du x^\mu A_\mu(u x) \big)$ has been omitted.
The coefficient in front of the r.h.s. of 
eqs. (\ref{phip}), (\ref{psigma})
can be written in terms of the light quark condensate
$<{\bar u} u>$ using the PCAC relation: 
$\displaystyle \mu_{\pi}= {m_\pi^2 \over m_u+m_d}
=-{2 \over f^2_\pi} <{\bar u} u>$. 

The PWF $\varphi_{\pi}(u)$ is associated with the leading twist two 
operator, $g_1(u)$ and $g_2(u)$ correspond to twist four operators, and $\varphi_P(u)$ and 
$\varphi_\sigma (u)$ to twist three ones. 
The function $\varphi_{3 \pi}$ is of twist three, while all the 
PWFs appearing in eqs.(\ref{gi}), (\ref{git}) are of twist four.
The PWFs $\varphi (x_i,\mu)$ ($\mu$ is the renormalization point) 
describe the distribution in longitudinal momenta inside the pion, the 
parameters $x_i$ ($\sum_i x_i=1$) 
representing the fractions of the longitudinal momentum carried 
by the quark, the antiquark and gluon.

The wave function normalizations immediately follow from the definitions
(\ref{phipi})-(\ref{git}):
$\int_0^1 du \; \varphi_\pi(u)=\int_0^1 du \; \varphi_\sigma(u)=1$,
$\int_0^1 du \; g_1(u)={\delta^2/12}$,
$\int {\cal D} \alpha_i \varphi_\bot(\alpha_i)=
\int {\cal D} \alpha_i \varphi_{\|}(\alpha_i)=0$,
$\int {\cal D} \alpha_i \tilde \varphi_\bot(\alpha_i)=-
\int {\cal D} \alpha_i \tilde \varphi_{\|}(\alpha_i)={\delta^2/3}$,
with the parameter $\delta$ defined by the matrix element: 
$<\pi(q)| {\bar d} g_s \tilde G_{\alpha \mu} \gamma^\alpha u |0>=
i \delta^2 f_\pi q_\mu$.

%%%%%%%%%%%%%%%%%%%%%%%%%%%%%%%%%%%%%%%%%%%%%%%%%%%%%%%%%%%%%%%%%%%%%%%%
\section{The LCQSR for the $\pi NN$ coupling}
\label{sec4}
%%%%%%%%%%%%%%%%%%%%%%%%%%%%%%%%%%%%%%%%%%%%%%%%%%%%%%%%%%%%%%%%%%%%%%%%
Expressing (\ref{quark}) with the pion wave functions, we arrive at:
\begin{eqnarray}\label{coordinate}\nonumber
\Pi (p_1, p_2, q) = \int d^4 x \int_0^1 du e^{i(p-uq) x} \{
{4f_\pi\over \pi^4 x^6} \{ [\varphi_\pi (u) +x^2g_1(u)] \gamma_5 {\hat q} 
-i\gamma_5 ({\hat x} -{x^2\over q\cdot x} {\hat q} g_2(u) \} &\\ \nonumber
-{f_\pi\over 9\pi^2x^4}\mu_\pi \langle {\bar q} q \rangle \varphi_\sigma (u) 
\gamma_5 (x^2 {\hat q}-q\cdot x {\hat x}) 
-{f_\pi\over 96\pi^2x^2}\mu_\pi m_0^2\langle {\bar q} q \rangle \varphi_\sigma (u) 
\gamma_5 (x^2 {\hat q}-q\cdot x {\hat x}) &\\  \nonumber
+{f_\pi\over 192\pi^2x^2} \langle g_s^2G^2 \rangle \varphi_\pi (u) 
\gamma_5 {\hat q} \} &\\ \nonumber 
-i\int d^4 x \int_0^1 du\int {\cal D} \alpha_i  e^{ip\cdot x-iq\cdot x (\alpha_1 +u\alpha_3)}
{f_{3\pi}\over 12\pi^2 x^2} \langle {\bar q} q \rangle \varphi_{3\pi} (\alpha_i )
[(1-2u) q^2 {\hat x} &\\
+2(1-2u) (q\cdot x) {\hat q} ] +\cdots  \;,
\end{eqnarray}
where $\mu_{\pi}=1.65$GeV, 
$f_{\pi}=132$MeV, $\langle {\bar q} q \rangle=-(225\mbox{MeV})^3$, 
$\langle {\bar q}g_s\sigma\cdot G q \rangle =m_0^2\langle {\bar q} q \rangle$, 
$m_0^2=0.8$GeV$^2$, $a=-(2\pi )^2 \langle {\bar q}q \rangle$.
We have collected the terms relevant for the tensor structure $\gamma_5 {\hat q}$ 
in (\ref{coordinate}) only.

In the following sections we will frequently use integration by parts 
to absorb the factors $(q\cdot x)$ and $1/(q\cdot x)$, which leads to the derivative 
and integration of PWFs. For example, 
\begin{equation}\label{derivative}
\int_0^1 {q\cdot x \over (x^2)^n }\varphi_\pi (u) e^{-iu q\cdot x} du =
-i\int { e^{-iu q\cdot x}\over (x^2)^n }\varphi^\prime_\pi (u)  du +
\varphi_\pi (u) e^{-iu q\cdot x}|_0^1 \; ,
\end{equation}
\begin{equation}\label{integration}
\int_0^1 {e^{-iu q\cdot x}\over  q\cdot x }g_2 (u)  du =
-i\int  e^{-iu q\cdot x} G_2 (u)  du -
G_2 (u) e^{-iu q\cdot x}|_0^1 \; ,
\end{equation}
where the functions $G_2(u)$ is defined as:
\begin{equation}
G_2 (u)=-\int_0^{u} g_2(u)du \; .
\end{equation}
Note the second term in (\ref{derivative}) and (\ref{integration}) vanishes due to 
$\varphi_\pi (u_0) =G_2 (u_0)=0$ at end points $u_0 =0, 1$.

We first finish Fourier transformation in (\ref{coordinate}). The formulas are: 
\begin{equation}\label{ft1}
\int {e^{ipx}\over (x^2)^n} d^D x = i (-1)^{n+1} 
{2^{D-2n}\pi^{D/2} \over (-p^2)^{D/2 -n}} {\Gamma (D/2 -n)\over \Gamma (n)} \; ,
\end{equation}
\begin{equation}\label{ft2}
\int {{\hat x}e^{ipx}\over (x^2)^n} d^D x =  (-1)^{n+1} 
{2^{D-2n+1}\pi^{D/2} \over (-p^2)^{D/2+1 -n}} {\Gamma (D/2+1 -n)\over \Gamma (n)}
{\hat p} \; .
\end{equation}

Making double Borel transformation with the variables $p_1^2$ and $p_2^2$
the single-pole terms in (\ref{pole}) are eliminated. The formula reads:
\begin{equation}\label{double}
{{\cal  B}_1}^{M_1^2}_{p_1^2} {{\cal  B}_2}^{M_2^2}_{p_2^2} 
{\Gamma (n)\over [ m^2 -(1-u)p_1^2-up^2_2]^n }=
(M^2)^{2-n} e^{-{m^2\over M^2}} \delta (u-u_0 ) \; .
\end{equation}

Subtracting the continuum contribution which is modeled by the 
dispersion integral in the region $s_1 , s_2\ge s_0$, we arrive at:
\begin{eqnarray}\label{quark0}\nonumber
\sqrt{2} m_N \lambda^2_N  g_{\pi NN}e^{-{ M_N^2\over M^2} } = &\\ \nonumber
e^{-{u_0(1-u_0)q^2\over M^2}} \{
+{f_\pi\over 2\pi^2} \varphi_\pi (u_0) M^6 f_2 ({s_0\over M^2})
-{4f_\pi\over \pi^2} [g_1(u_0)+G_2 (u_0) ] M^4 f_1 ({s_0\over M^2})  
&\\ \nonumber
+{f_\pi\over \pi^2} u_0 g_2 (u_0) M^4 f_1 ({s_0\over M^2})  
+{f_\pi\over 9\pi^2} a\mu_\pi [\varphi_\sigma (u_0) 
+{u_0\over 2}\varphi_\sigma^\prime (u_0)] M^2 f_0 ({s_0\over M^2}) 
&\\ \nonumber
+{f_\pi\over 48\pi^2} \langle g_s^2 G^2\rangle 
\varphi_\pi (u_0) M^2 f_0 ({s_0\over M^2})
-{f_\pi\over 48\pi^2} am_0^2\mu_\pi u_0 \varphi_\sigma^\prime (u_0) 
&\\ 
+{f_{3\pi}\over 6\pi^2}a M^2 f_0 ({s_0\over M^2}) [I_1^G (u_0) +I_2^G (u_0) -2I_3^G (u_0) ]
-{f_{3\pi}\over 6\pi^2}a q^2 u_0 I_4^G (u_0) 
\}&\; ,
\end{eqnarray}
where
$f_n(x)=1-e^{-x}\sum\limits_{k=0}^{n}{x^k\over k!}$ is the factor used 
to subtract the continuum, $s_0$ is the continuum threshold.
$u_0={M^2_1 \over M^2_1 + M^2_2}$, 
$M^2\equiv {M^2_1M^2_2\over M^2_1+M^2_2}$, 
$M^2_1$, $M^2_2$ are the Borel parameters, 
and $\varphi_\pi^\prime (u_0)  ={d\varphi_\pi (u)\over du}|_{u=u_0}$ etc.

The functions $I^G_i (u_0)$ are defined as:
\begin{equation}
I^G_1 (u_0) =\int_0^{u_0} d\alpha_1 
{\varphi_{3\pi}(\alpha_1, 1-u_0, u_0-\alpha_1) \over u_0-\alpha_1}
\; ,
\end{equation}
\begin{equation}
I^G_2 (u_0) =\int_0^{1-u_0} d\alpha_2 
{\varphi_{3\pi}(u_0, 1-u_0, \alpha_2, 1-u_0-\alpha_2) \over 1-u_0-\alpha_2}
\; ,
\end{equation}
\begin{equation}
I^G_3 (u_0) =\int_0^{u_0} d\alpha_1 \int_0^{1-u_0} d\alpha_2
{\varphi_{3\pi}(\alpha_1, \alpha_2, 1-\alpha_1-\alpha_2) \over (1-\alpha_1-\alpha_2)^2}
\; ,
\end{equation}
\begin{equation}
I^G_4 (u_0) =\int_0^{u_0} d\alpha_1 \int_0^{1-u_0} d\alpha_2
{\varphi_{3\pi}(\alpha_1, \alpha_2, 1-\alpha_1-\alpha_2) \over (1-\alpha_1-\alpha_2)^2}
(1-2u_0+\alpha_1-\alpha_2) u_0
\; .
\end{equation}

The twist four PWFs $\varphi_\bot(\alpha_i)$, $ \varphi_{\|}(\alpha_i)$,
$ \tilde \varphi_\bot(\alpha_i)$ and $ \tilde \varphi_{\|}(\alpha_i)$
do not contribute to the chiral odd tensor structures, which was 
first observed in \cite{bely-z}. 
Moreover our sum rule (\ref{quark0}) is symmetric with the Borel parameters 
$M_1^2$ and $M_2^2$. So it's natural to adopt
$M_1^2=M_2^2=2M^2$, $u_0 ={1\over 2}$.
We shall work in the physical limit $q^2=m_\pi^2 \to 0$ and discard
the terms with the factor $q^2$ in (\ref{quark0}).

The various parameters which we adopt are 
$a=0.55\, \mbox{GeV}^3$, $\langle g_s^2 G^2\rangle=0.48\, \mbox{GeV}^4$, 
$m_0^2 =0.8\, \mbox{GeV}^2$ at the scale $\mu =1$GeV, 
$s_0=2.25$GeV$^2$, $m_N=0.938$GeV, 
$\lambda_N =0.026$GeV$^3$ \cite{IOFFE}.  

The working interval for analyzing the QCD sum rule (\ref{quark0}) 
is $0.9\mbox{GeV}^2 \leq M_B^2\leq 1.8\mbox{GeV}^2$, a standard choice 
for analyzing the various QCD sum rules associated with the nucleon. 
In order to diminish the uncertainty due to $\lambda_N$, we shall 
divide (\ref{quark0}) by the famous Ioffe's mass sum rule for the nucleon:
\begin{equation}\label{mass}
32\pi^4 \lambda_N^2 e^{-{ M_N^2\over M^2} }
=M^6 f_2 ({s_0\over M^2})+{b\over 4}M^2 f_0 ({s_0\over M^2})
+{4\over 3}a^2 -{a^2m_0^2\over 3M^2} \; .
\end{equation}

The resulting sum rule depends on the PWFs, the integrals and derivatives of them 
at the point $u_0={1\over 2}$. Since $\delta^2$ is numerically small, 
the uncertainty due to the integral term $G_2(u_0)$ is insignificant.
There are many discussions about the leading twist PWF $\varphi_\pi (u)$ in literature 
\cite{cz,dziem,bely95,ht,halperin,rad,johnson}. 
For example, at $u_0={1\over2}$ the values of various model functions are: 
$\varphi_\pi(u_0)=1.22$ \cite{bely95}, 
$1.273$ \cite{rad}, $1.25$ \cite{halperin}, 
$1.71$ \cite{ht}, $1.35$ \cite{johnson} and $1.5$ 
for the asymptotic PWF respectively.

Knowledge of the PWFs involved with the gluon field is still very limited. Only the 
very lowest a few moments of these PWFs were calculated with significant errors using 
the method of QSR, which in turn were used to determine the detailed forms of the 
functions. Such an approach is sometimes misleading since   
there are many wave functions satisfying the same constraints from moments.

In order to illustrate this point more clearly we use the determination of 
$\varphi_\pi (u)$ as an example. The model wave function 
for $\varphi_\pi (u)$ based on the QCD sum rule approach 
was given in \cite{bely95} as:
\begin{equation}\label{wf-0}
\varphi_\pi (u,\mu) =6u{\bar u}\left(
1+a_2(\mu){3\over 2}[5(u-{\bar u})^2-1]+a_4(\mu) {15\over 8} 
[21(u-{\bar u})^4-14(u-{\bar u})^2+1]\right),
\end{equation}
where ${\bar u}=1-u$.
Yet the authors pointed out themselves that the oscillations of (\ref{wf-0}) 
around $u_0={1\over 2}$ is unphysical,
which is due to the truncation of the series and keeping only the first a few 
terms when $\varphi_\pi (u)$ is expanded over Gegenbauer polynomials. In 
\cite{halperin} it was stressed that: (1) the expansion over Gegenbauer polynomials 
converges very slowly as can be seen from the large value of $a_2$ and $a_4$; 
(2) any oscillating wave function is not physical, since no detached scale is seen
to govern such oscillations. Recently Mikhailov and Radyushkin \cite{rad} 
reanalyzed the QCD sum rules for $\varphi_\pi (u)$ taking into 
account the non-locality of the condensates. They suggested the following 
wave function:
\begin{equation}\label{wf-a}
\varphi_\pi (u) ={8\over \pi}\sqrt{u{\bar u}}\;.
\end{equation}
Halperin suggested the following form \cite{halperin}:
\begin{equation}\label{wf-b}
\varphi_\pi (u) =N exp\left( -{m^2\over 8\beta^2 u {\bar u}} \right)
[\mu^2+\mu {\tilde \mu}+({\tilde\mu}^2-2)u{\bar u}]
\;,
\end{equation}
where $N=4.53$, $\mu={m\over 2\beta}$, ${\tilde\mu}={1\over 2\beta}\left(
{1\over 4}m_\pi +{3\over 4}m_\rho\right)={{\tilde m}\over 2\beta}$ with 
$m=330$MeV, ${\tilde m}=620$MeV, and $\beta=320$MeV.
Note all the above three forms of PWFs are rather close to the asymptotic form 
\begin{equation}\label{wf-d}
\varphi^{\mbox{asym}}_\pi (u) =6u{\bar u}\; ,
\end{equation}
which are in strong contrast with the original double humped Chernyak-Zhitnitsky 
form keeping the lowest two orders of the expansion \cite{cz}:
\begin{equation}\label{wf-e}
\varphi^{\mbox{asym}}_\pi (u) =30u{\bar u}(1-4u{\bar u})\; .
\end{equation}
In other words, after summing the whole series the physical PWF  
can not deviate too much from the asymptotic form. 
Based on the above consideration we use the asymptotic forms for 
the PWFs involved with gluons as in \cite{zhu2}.

For the other PWFs we use the results given in \cite{bely95} 
since they are relatively
better known. $\varphi_P(u_0)=1.19$, $\varphi_\sigma(u_0)=1.47$,  
$g_1(u_0)=0.022 $GeV$^2$, $g_2 (u_0) =0$ and 
$G_2(u_0)=0.02$GeV$^2$ at $u_0 ={1\over 2}$
and $\mu =1$GeV. Their first derivatives satisfy:
$\varphi^\prime_\pi(u_0)=\varphi^\prime_P(u_0) =\varphi^\prime_\sigma(u_0)=
g^\prime_1(u_0)=G_2^\prime(u_0)=0$. 

The dependence on the Borel parameter $M^2$ of $g_{\pi NN}$ 
are shown in FIG 2 with $s_0=2.25$ GeV$^2$ using different values 
of $\varphi_\pi(u_0)$. From top to bottom the curves 
corresponds to $\varphi_\pi(u_0)=1.6, 1.5, 1.4$ respectively. 
The final sum rule is very stable in the working region of the Borel 
parameter $M^2$ and sensitive to the value of $\varphi_\pi ({1\over 2})$.
Experimentally the $\pi NN$ coupling constant has been extracted very precisely:
$g_{\pi NN}=13.5$ \cite{nijimegen}. Using this value as the input we obtain:
\begin{equation}\label{num}
\varphi_\pi(u_0, \mu =1\mbox{GeV})=1.5\pm 0.2 \; ,
\end{equation}
which is very close to the asymptotic PWF. 

It is interesting to notice that
the QSR derived from the chiral odd structure ${\hat q}\gamma_5$ yielded
$\varphi_\pi(u_0)=1.6$ while the QSR from the chiral even structure 
${\hat q}{\hat p}\gamma_5$ led to a smaller value \cite{bely-z}.
After averaging the two values the authors got $\varphi_\pi(u_0)=1.2\pm 0.3$.
As pointed out in Section \ref{sec2}, generally the QSR derived from chiral
odd structure is more reliable. Therefore, such an averaging may not be 
very suitable here. 

Very recently Belyaev and Johnson investigated the relation between 
light-front quark model and QCD \cite{johnson}. They found 
additional support that the two point PWF $\varphi_\pi (u)$ 
attains the asymptotic form. Numerically $\varphi_\pi (u)$ at 
$u={1\over 2}$ agrees with the asymptotic PWF within $10\%$, 
which was consistent with our result. It appears that almost all 
the phenomenological analyses share the common feature, i.e., 
that the PWF starts to approach the asymptotic form more or less  
at the low scale $\mu =1$GeV already. This point deserves further 
investigation.

%%%%%%%%%%%%%%%%%%%%%%%%%%%%%%%%%%%%%%%%%%%%%%%%%%%%%%%%%%%%%%%%%%%%%%%%
\section{The LCQSR for the $\pi NN(1535)$ coupling}
\label{sec5}
%%%%%%%%%%%%%%%%%%%%%%%%%%%%%%%%%%%%%%%%%%%%%%%%%%%%%%%%%%%%%%%%%%%%%%%%
Quantum Chromodynamics (QCD) is asymptotically free and its high energy 
behavior has been tested to one-loop accuracy. On the other hand, the 
low-energy behavior has become a very active research field in the 
past years. Various hadronic resonances act as suitable labs for 
exploring the nonperturbative QCD.  
The inner structure of nucleon and mesons and their interactions is of central
importance in nuclear and particle physics. 
Internationally there are a number of experimental  
collaborations, like TJNAL (former CEBAF), 
COSY, ELSA (Bonn), MAMI (Mainz) and Spring8 (Japan), 
which will extensively study the excitation of higher 
nucleon resonances.

Among various baryon resonances, negative parity resonance $N^\ast (1535)$
is particularly interesting, which dominates the $\eta$ meson photo- or 
electro-production on a nucleon. The branching ratio for the decay 
$N^\ast \to \eta N$ is comparable with that for $N^\ast \to \pi N$. 
Considering the phase space difference and using the experimental decay 
width of $N^\ast$ \cite{review}, we get $g_{\eta N N^\ast}=2$ 
and $g_{\pi N N^\ast}=0.7$. 
The latter value is in strong contrast with the pion nucleon coupling
$g_{\pi NN}=13.4$. Thus arises the question: why is the coupling 
$g_{\pi N N^\ast}$ so small compared with $g_{\pi NN}$?

Whether the coupling $g_{\pi N N^\ast}$ is strongly suppressed is under 
heated debate in literature. In a recent extended coupled channel 
analysis of $\pi N$ scattering, the J$\ddot{u}$lich group used 
$g_{\eta N N^\ast}=1.94$ and $g_{\pi N N^\ast}<0.12$ \cite{julich}. 
Jido, Oka and Hosaka suggested in a recent letter that $\pi NN^\ast$ 
coupling is strongly suppressed as a consequence of chiral symmetry \cite{jido}. 
Their argument was based on a pion-nucleon correlator of two baryon 
interpolating fields. The chiral transformation properties of their 
interpolating fields then imply that the correlator is purely proportional 
to the tensor structure $\gamma_5$, with no piece of the form 
${\hat p}\gamma_5$, which is the relevant structure for $\pi N N^\ast$ coupling.
Based on the above observation they claimed that the coupling $\pi N N^\ast$
vanishes. The above argument was criticized by Birse \cite{birse}. 
He pointed out that the ${\hat p}\gamma_5$ piece of the correlator is a sum 
of all possible pion-baryon couplings that can contribute. Hence the absence
of a ${\hat p}\gamma_5$ piece is a statement about the particular combination
of the pion-baryon coupling and the subtraction terms that correspond to the 
chosen interpolating fields. It does not imply that the physical $\pi NN^\ast$
coupling is suppressed. 
With an interpolating field with covariant derivative 
for the $N^\ast$ Kim and Lee used QCD sum rules to estimate 
$g_{\pi NN^\ast} \sim 1.5$ \cite{lee}. 
But in their analysis the continuum and excited states contribution 
is poorly subtracted and the numerical results depend strongly 
on the value of the quark gluon mix condensate 
$\langle {\bar q}g_s \sigma\cdot G q\rangle$, 
which renders their conclusion less convincing.

In this section we shall employ the LCQSR to calculate $\pi NN^\ast$ coupling. 
The continuum and excited states contribution is subtracted 
rather cleanly within our approach through double Borel transformation.

We start with the two point function 
\begin{equation}
\Pi (p_1,p_2,q) = \int d^4 x e^{ip x} 
\left \langle 0 \vert {\cal T}
\eta_p (x;s)  {\bar{\eta_n}} (0;t) \vert \pi^+ (q) \right \rangle
\label{eq1}
\end{equation}
with $p_1 =p$, $p_2 = p-q$ and 
the general nucleon interpolating field without derivatives 
which couples to both positive and negative parity nucleon resonances
\begin{equation}\label{cur2}
\eta_p (x;s) = \epsilon_{abc} \{
 \left [ u^a (x) {\cal C} d^b (x) \right ] \gamma_5  u^c (x) +
s \left [ u^a (x) {\cal C}\gamma_5 d^b (x) \right ]   u^c (x)
\}
\label{eq2}
\end{equation}
where $s$ is the mixing parameter and ${\bar{\eta_p}}={\eta_p}^\dag \gamma_0$.
Note we use a slightly different nucleon interpolating field (\ref{cur2}) 
in this section. The current $\eta_p (x;s=-1)$ is one half of 
the one defined in \ref{cur1}. Both of them are called Ioffe's current.
The later form appears more in literature. The Ioffe's current
couples strongly to the positive parity nucleon 
\cite{IOFFE,espriu}, while it was found that the current $\eta_p (x;s=0.8)$ 
is optimized for negative parity nucleons and couples strongly to 
$N(1535)$ \cite{jido2}. 

It is important to note that the diagonal transitions like $N\to N$, $N^\ast \to N^\ast$
does not contribute to the tensor structure ${\hat p} \gamma_5$. 
In other words, the function $F_2$ involves solely with the process 
$N^\ast \to N$ and corresponding continuum contribution. 
Based on the above observation we shall focus 
on the chiral odd tensor structure ${\hat p} \gamma_5$ in this section.

The $\pi N N(1535)$ coupling constant $g_{\pi NN^\ast}$ is defined as:
\begin{equation}
\label{def}
{\cal L}_{\pi NN^\ast} =  g_{\pi NN^\ast} {\bar N}
{\bf \tau \cdot \pi} N + h.c.
\end{equation}

At the phenomenological level the eq.(\ref{eq1}) can be expressed as:
\begin{equation}
\Pi (p_1,p_2,q)  = -(m_N +m_{N^\ast})   g_{\pi NN^\ast}
\{  {\lambda_N (s)\lambda_{N^\ast} (t)\over{ (p_1 ^2 - M_N^2) (p_2 ^2 - M_{N^\ast}^2) }} 
-{\lambda_{N^\ast} (s)\lambda_N (t)\over{ (p_1 ^2 - M_{N^\ast}^2) (p_2 ^2 - M_N^2) }} 
\} i {\hat p} \gamma_5 +\cdots
\label{phen2}
\end{equation}
where we write the structure ${\hat p} \gamma_5$ explicitly only
and the continuum contribution is denoted by the ellipse. 
$\lambda_N (s)$ is the overlapping amplitude of 
the interpolating current $\eta_N (x)$ with the nucleon state
\begin{equation}
\left \langle 0 \vert \eta (0;s) \vert N (p) \right \rangle 
= \lambda_N(s)  u_N (p)
\end{equation}

We neglect the four particle component of the PWF	
and express (\ref{eq1}) with the PWFs.
After Fourier transformation and making double Borel transformation 
with the variables $p_1^2$ and $p_2^2$
the single-pole terms are eliminated and finally we arrive at:
\begin{eqnarray}\label{quark1}\nonumber
(m_N +m_{N^\ast}) [ \lambda_N (s)\lambda_{N^\ast} (t) -
\lambda_{N^\ast} (s)\lambda_N (t)] g_{\pi NN^\ast}
e^{-({ M_N^2\over M_1^2}+{ M_{N^\ast}^2\over M_2^2}) } = &\\ \nonumber
-{f_\pi\over 192\pi^2} (1+s)(1+t)
\varphi_\pi^\prime (u_0) M^6 f_2 ({s_1\over M^2})
+{f_\pi\over 16\pi^2} (1+s)(1+t)
g_1^\prime (u_0) M^4 f_1 ({s_1\over M^2})  &\\ \nonumber
-{f_\pi\over 32\pi^2} [7(1+st)+3(s+t)]
g_2 (u_0) M^4 f_1 ({s_1\over M^2}) 
+{f_\pi\over 288\pi^2} (1+s+t-3st)a\mu_\pi
\varphi_\sigma^\prime (u_0) M^2 f_0 ({s_1\over M^2}) &\\ \nonumber
+{f_\pi\over 4608\pi^2} (1+s)(1+t)\langle g_s^2 G^2\rangle 
\varphi_\pi^\prime (u_0) M^2 f_0 ({s_1\over M^2})
-{f_\pi\over 6912\pi^2} (5+3s+3t-11st)am_0^2\mu_\pi
\varphi_\sigma^\prime (u_0) &\\ \nonumber
+{f_\pi\over 192\pi^2} (s-t)am_0^2\mu_\pi \varphi_P (u_0)
-{f_\pi\over 2304\pi^2} [19(1+st)+7(s+t)] \langle g_s^2 G^2\rangle g_2 (u_0) &\\
-{f_\pi\over 1152\pi^2} (1+s)(1+t)\langle g_s^2 G^2\rangle g_1^\prime (u_0)&\; .
\end{eqnarray}

The twist three PWF $\varphi_{3 \pi}$
appears in the combination $f_{3\pi} \varphi_{3 \pi} \langle {\bar q} q\rangle q^2$.
In the physical limit $q^2=m_\pi^2 \to 0$, its contribution is negligible, 
which is in contrast with the sum rule for $g_{\pi NN}$ (\ref{quark0}), where 
the twist three PWF $\varphi_{3 \pi}$ plays an important role. 

It is a common practice to adopt the Ioffe current for the ground-state 
nucleon \cite{IOFFE}, i.e., letting $s=-1$. Now the equation (\ref{quark1}) has 
a simple form:
\begin{eqnarray}\label{quark2}\nonumber
(m_N +m_{N^\ast}) [ \lambda_N (s)\lambda_{N^\ast} (t) -
\lambda_{N^\ast} (s)\lambda_N (t)] g_{\pi NN^\ast}
e^{-({ M_N^2\over M_1^2}+{ M_{N^\ast}^2\over M_2^2}) } = &\\ \nonumber
-{f_\pi\over 8\pi^2} (1-t) g_2 (u_0) M^4 f_1 ({s_1\over M^2}) 
+{f_\pi\over 72\pi^2} t a\mu_\pi
\varphi_\sigma^\prime (u_0) M^2 f_0 ({s_1\over M^2}) &\\ \nonumber
-{f_\pi\over 3456\pi^2} (1+7t)am_0^2\mu_\pi \varphi_\sigma^\prime (u_0) 
-{f_\pi\over 192\pi^2} (1+t)am_0^2\mu_\pi \varphi_P (u_0) &\\
-{f_\pi\over 192\pi^2} (1-t)\langle g_s^2 G^2\rangle g_2 (u_0) & \; ,
\end{eqnarray}
where $t=0.8$.

The various parameters are 
$s_1=3.2$GeV$^2$, $m_{N^\ast}=1.535$GeV, 
$\lambda_N (s=-1)=0.013$GeV$^3$, $\lambda_{N^\ast} (s=0.8)=0.27$GeV$^3$ 
with the formulas in \cite{jido2}. Moreover 
$\lambda_N (s=-1) \gg \lambda_N (t=0.8)$, 
$\lambda_{N^\ast} (t=0.8) \gg \lambda_{N^\ast} (s=-1)$, so it is 
reasonable to discard the term $\lambda_N (t=0.8) \lambda_{N^\ast} (s=-1)$ 
in the eq. (\ref{quark2}). We use the model PWFs presented in \cite{bely95}
to make the numerical analysis.

Our sum rule (\ref{quark2}) is asymmetric with the Borel parameters 
$M_1^2$ and $M_2^2$ due to the significant mass difference of $N$ and $N(1535)$.
It is natural to let $M_1^2=2m_N^2 \beta$, $M_2^2=2m_{N^\ast}^2 \beta$,
where $\beta$ is a scale factor ranging from $1.0$ to $2.0$. In this case 
we have $M^2 =1.28\beta$GeV$^2$, $u_0 =0.27$, $g_2(u_0)=-0.03$GeV$^2$, 
$\varphi_P (u_0)=0.66$ and $\varphi^\prime_\sigma (u_0)=2.63$ at the 
scale $\mu =1$GeV. 

The sum rule for $g_{\pi N N^\ast}$ is very stable with reasonable variations 
of $s_1$ and $M^2$ as can be seen in FIG 3. Finally we have 
\begin{equation}\label{final}
g_{\pi N N^\ast}=(-) (0.08 \pm 0.03) \; .
\end{equation}
Our result is consistent with the conclusions in Refs. \cite{jido,julich}.

We have included the uncertainty due to the variation of the continuum 
threshold and the Borel parameter $\beta$ in (\ref{final}). 
In other words, only the errors arising from numerical analysis of 
the sum rule (\ref{quark2}) are considered. Other sources of 
uncertainty include: (1) the truncation of OPE on the light cone and keeping 
only the lowest twist operators. For example the four particle 
component of PWF is discarded explicitly; (2) the inherent 
uncertainty due to the detailed shape of PWFs in different models; 
(3) throwing away the term $\lambda_N (t=0.8) \lambda_{N^\ast} (s=-1)$ 
in the eq. (\ref{quark2}); (4) the continuum model etc. 
In the present case the major error comes from the 
uncertainty of PWFs since our final sum rule 
(\ref{quark2}) depends both on the value of PWFs and their derivatives
at $u_0=0.27$. Luckily around $u_0=0.27$ different model PWFs 
have a shape roughly consistent with each other. So a more 
conservative estimate is to enlarge the error by a factor of two.
Now we have 
\begin{equation}\label{final2}
g_{\pi N N^\ast}=(-) (0.08 \pm 0.06) \; .
\end{equation}
We want to point out that one should not be too serious about 
the specific number. What's important is 
the fact that $|g_{\pi N N^\ast}| \ll |g_{\pi N N}|$. 
%%%%%%%%%%%%%%%%%%%%%%%%%%%%%%%%%%%%%%%%%%%%%%%%%%%%%%%%%%%%%%%%%%%%%%%%
\section{Discussion}
\label{sec6}
%%%%%%%%%%%%%%%%%%%%%%%%%%%%%%%%%%%%%%%%%%%%%%%%%%%%%%%%%%%%%%%%%%%%%%%%
In summary we have constrained the parameter $\varphi_\pi ({1\over 2})$ 
using the experimentally precisely known $g_{\pi NN}$ using LCQSR. We also 
have also calculated the $\pi N N^\ast$ coupling. 
The continuum and the excited states contribution is subtracted rather 
cleanly through the double Borel transformation in both cases.
Our result shows explicitly the suppression of $g_{\pi N N^\ast}$.

\vspace{0.8cm} {\it Acknowledgments:\/} S.-L. Zhu  was in part 
supported by the National Postdoctoral Science Foundation of China
and National Natural Science Foundation of China. Y.D. 
was supported by the National Natural Science Foundation of China.
W-Y. Hwang was supported by the National Science Council of ROC 
(Grant No. NSC86-2112-M002-010Y).

\newpage
{\bf Figure Captions}
\vspace{2ex}
\begin{center}
\begin{minipage}{130mm}
{\sf FIG 1.} \small{
The relevant feynman diagrams for the derivation of the LCQSR for 
$\pi NN$ and $\pi NN(1535)$ coupling. The squares 
denote the pion wave function (PWF). The broken solid line, broken curly line
and a broken solid line with a curly line attached in the middle
stands for the quark condensate, gluon condensate 
and quark gluon mixed condensate respectively.  }
\end{minipage}
\end{center}
\begin{center}
\begin{minipage}{130mm}
{\sf FIG 2.} \small{The sum rule for $g_{\pi NN}$ as a functions of the Borel parameter
$M^2$ for different $\varphi_\pi ({1\over 2})$ with the continuum threshold 
$s_0 =2.25$GeV$^2$. From bottom to top the curves correspond to 
$\varphi_\pi ({1\over 2})=1.6, 1.5, 1.4$ respectively.
}
\end{minipage}
\end{center}
\begin{center}
\begin{minipage}{130mm}
{\sf FIG 3.} \small{The sum rule for $g_{\pi NN^\ast}$ as a function of the scale 
parameter $\beta$ with the continuum threshold $s_1=3.4, 3.2, 3.0$ GeV$^2$ 
using the PWFs in \cite{bely95}. }
\end{minipage}
\end{center}


\bigskip
\vspace{1.cm}

\begin{thebibliography}{99}
\bibitem{SVZ}M.A. Shifman, A.I. Vainshtein, and V.I. Zakharov, Nucl. Phys. B
{\bf 147}, 385, 448, 519 (1979).

\bibitem{RRY}L.I. Reinders, H.R. Rubinstein, and S. Yazaki, Nucl. Phys. 
B{\bf 196}, 125 (1982);
L.I. Reinders, H.R. Rubinstein, and S. Yazaki, Phys. Rep.{\bf 127}, 1 (1985).

\bibitem{IOFFE}B.L. Ioffe, Nucl. Phys. B{\bf 188}, 317 (1981); [E] 
B{\bf 191}, 591 (1981);
V.M. Belyaev and B.L. Ioffe, Zh. Eksp. Teor. Fiz. {\bf 83}, 876 (1982)
[Sov. Phys. JETP{\bf 56}, 493 (1982)];
B.L. Ioffe and A.V.Smilga, Phys. Lett. B{\bf 114}, 353 (1982);
B.L. Ioffe and A.V. Smilga, Nucl. Phys. B{\bf 232}, 109 (1984).

\bibitem{BALIT}I.I.Balitsky and A.V.Yung, Phys. Lett. B{\bf 129}, 328 (1983).

\bibitem{bely95}
I. I. Balitsky, V. M. Braun and A. V. Kolesnichenko, Nucl. Phys. B 
{\bf 312}, 509 (1989);\\
V. M. Braun and I. E. Filyanov, Z. Phys. C {\bf 48}, 239 (1990);\\
V. M. Belyaev, V. M. Braun, A. Khodjamirian and R. R\"uckl, Phys. Rev. {\bf D 51}, 
6177 (1995).

\bibitem{bely-z}
V. M. Braun and I. E. Filyanov, Z. Phys. C {\bf 44}, 157 (1989).

\bibitem{jetp}V. M. Belyaev and Ya. I. Kogan, JETP Lett. {\bf 37}, 730 (1983).

\bibitem{ioffe98}B. L. Ioffe and A. G. Oganesian, Phys. Rev. D {\bf 57}, 
R6590 (1998).

\bibitem{zhu1}
Yuan-Ben Dai and Shi-Lin Zhu, hep-ph/9802227, Euro. Phys. J. C (in press). 

\bibitem{zhu3}
Yuan-Ben Dai and Shi-Lin Zhu, hep-ph/9802224, Phys. Rev. D (in press).

\bibitem{zhu2}
Shi-Lin Zhu and Yuan-Ben Dai, hep-ph/9802226, Phys. Lett. B {\bf 429}, 72 (1998).

\bibitem{zhu4}
Shi-Lin Zhu and Yuan-Ben Dai, hep-ph/9802225, Phys. Rev. D (in press).

\bibitem{bagan98}
E. Bagan, P. Ball and V. M. Braun, Phys. Lett. B {\bf 417}, 154 (1998).

\bibitem{zsl}W-Y. P. Hwang, Ze-sen Yang, Y.S. Zhong, Z.N. Zhou and 
Shi-Lin Zhu, Phys. Rev. {\bf C57}, 61 (1998).

\bibitem{nijimegen}J.J. De Swart, M.C.M. Rentmeester and R.G.E. Timmermans, 
Nucl-th/9802084, THEF-NYM-97.01, KVI 1318.

\bibitem{RRY2}L.J.Reinders, H.Rubinstein and S.Yazaki, Nucl.Phys. B {\bf 213},
109 (1983).

\bibitem{Rei}L.J.Reinders, Act.Phys.Pol. {\bf 15}, 329 (1984).

\bibitem{NP}S.Narison and N.Paver, Phys.Lett. B {\bf 135}, 159 (1984).

\bibitem{SH}H.Shiomi and T.Hatsuda, Nucl. Phys. A {\bf 594}, 294 (1995).

\bibitem{maltman}K. Maltman, Phys. Rev. C {\bf 57}, 69 (1998).

\bibitem{zsl-mag}Shi-Lin Zhu, W-Y.P. Hwang and Ze-sen Yang, Phys. Rev. 
D {\bf 57}, 1527 (1998).

\bibitem{cz}V. L. Chernyak and A. R. Zhitnitsky, Phys. Rep. {\bf 112}, 173(1984).

\bibitem{dziem}Z. Dziembowsky and L. Mankiewicz, 
Phys. Rev. Lett. {\bf 58}, 2175(1987).

\bibitem{ht}S. J. Brodsky, T. Huang, and G. P. Lepage, in {\sl Quarks and 
Nuclear Forces}, edited by D. Fries and B. Zeitnitz, Springer Tracts
in Modern Physics, Vol. {\bf 100} (Springer-Verlag, New York, 1982).

\bibitem{halperin}I. Halperin, Z. Phys. C{\bf 56}, 615(1992).

\bibitem{rad}S. V. Mikhailov and A. V. Radyushkin, 
Phys. Rev. D {\bf 45}, 1754(1992).

\bibitem{johnson}V. M. Belyaev and M. Johnson, Phys. Lett. B {\bf 423}, 
379 (1998).

\bibitem{review}  Particle Data Group, R. M. Barnett $et.~ al.$, Phys. Rev. 
{\bf D 54}, 1 (1996).

\bibitem{julich}C. Sch$\ddot{u}$tz, J. Haidenbauer, J. Speth and J.W. Durso, 
Phys. Rev. C {\bf 57}, 1464 (1998).

\bibitem{jido}D. Jido, M. Oka and A. Hosaka, Phys. Rev. Lett. {\bf 80}, 448 (1998).

\bibitem{birse}M.C. Birse, hep-ph/9801413, MC/TH98/01.

\bibitem{lee}H. Kim and S.-H. Lee, Phys. Rev. D {\bf 56}, 4278 (1997).

\bibitem{espriu}D. Espriu, P. Pascual and R. Tarrach, Nucl. Phys. B {\bf 214}, 
285 (1983).

\bibitem{jido2}D. Jido, N. Kodama and M. Oka, Phys. Rev. D {\bf 54}, 4532 (1996).

\end{thebibliography}
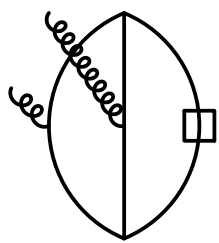
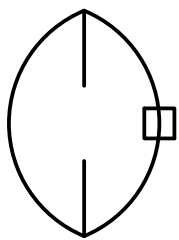
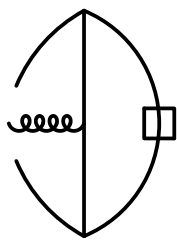
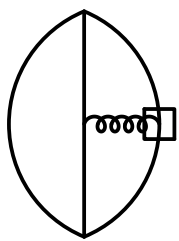
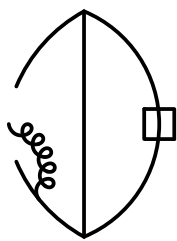
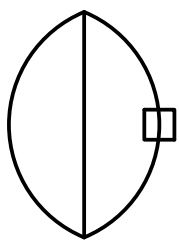
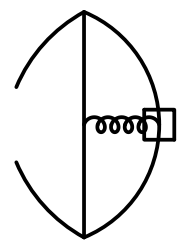
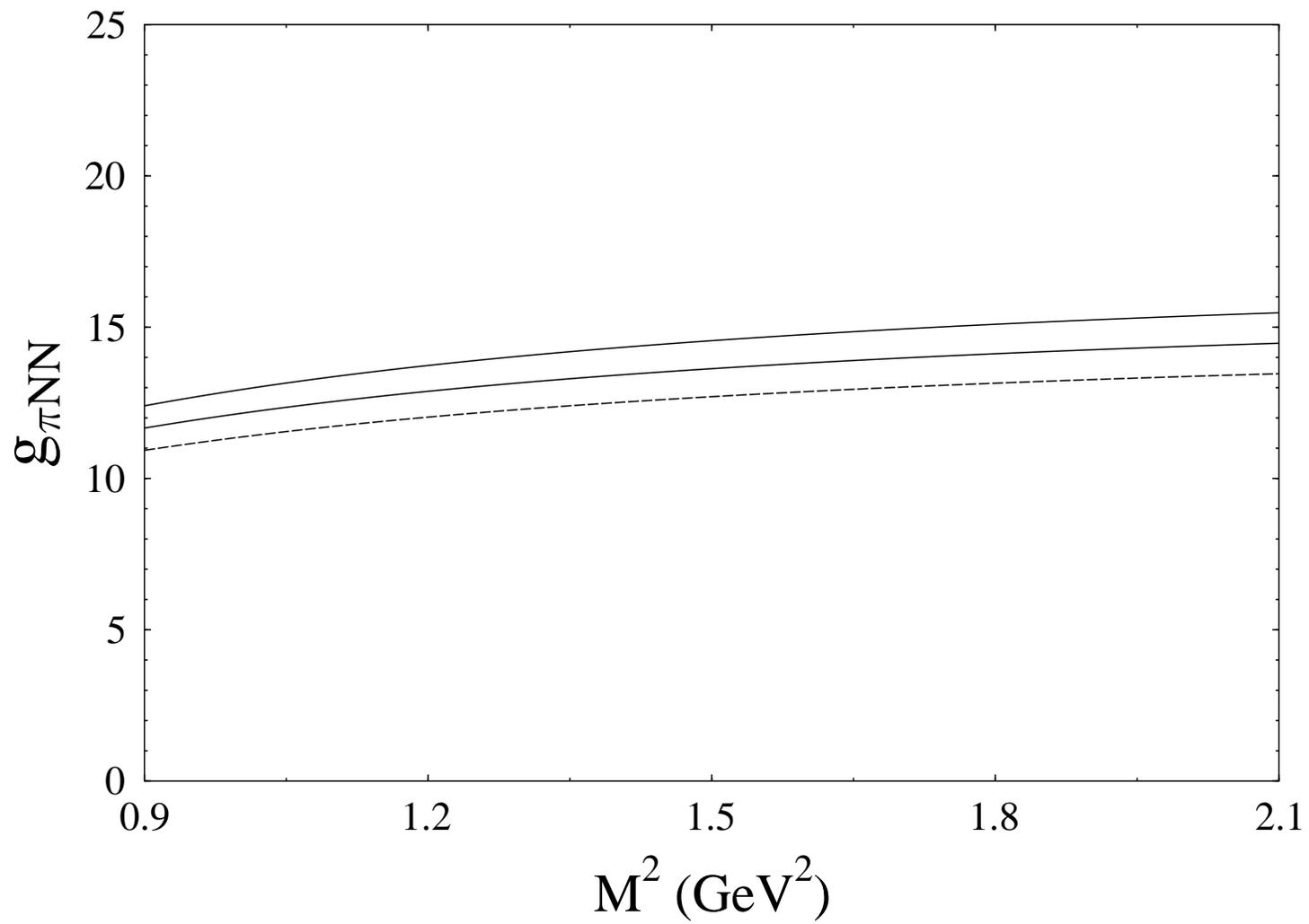
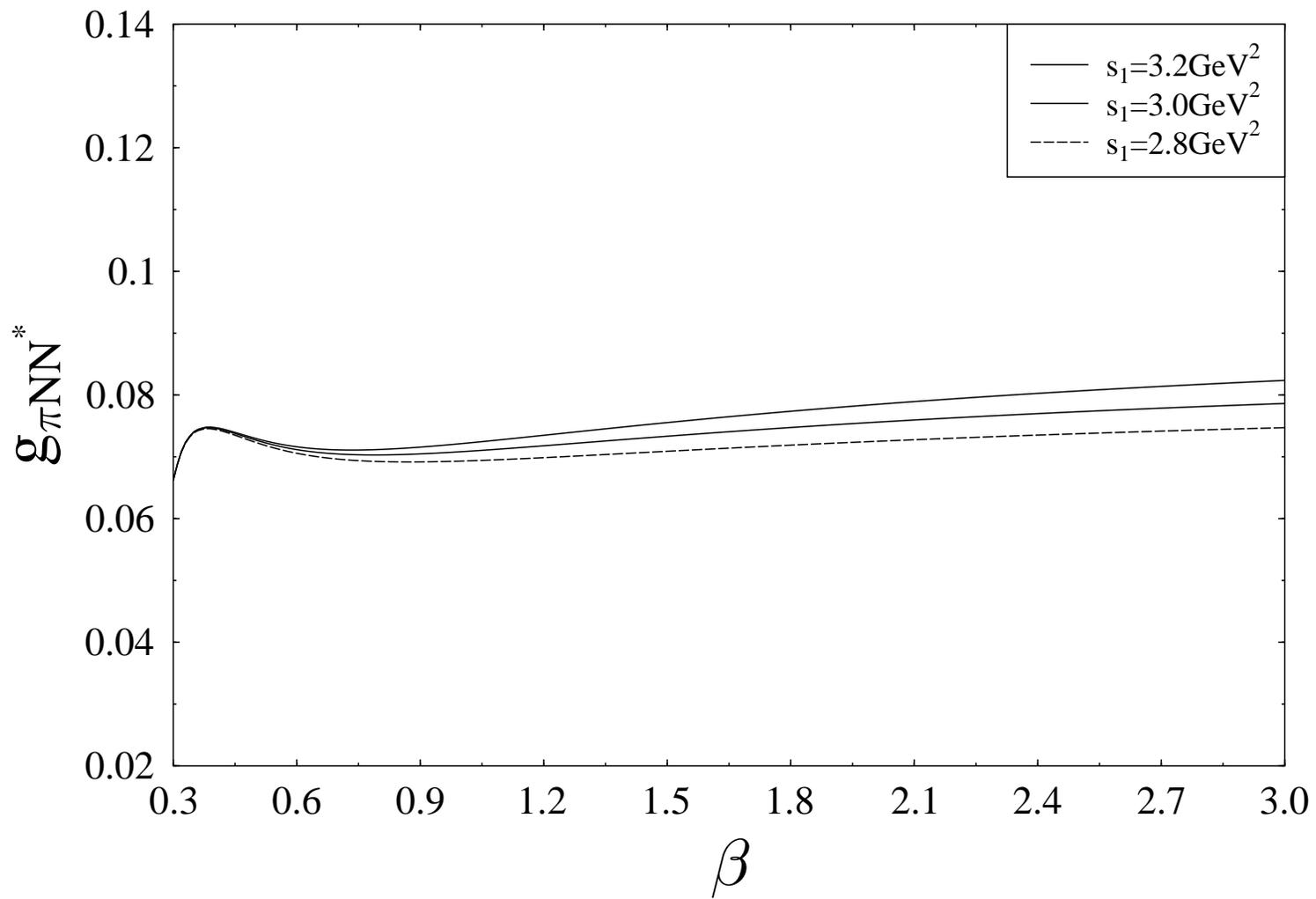
\end{document}